\documentclass[prb, twocolumn, amsmath, amssymb, showpacs]{revtex4-1}

\usepackage{CJK}
\usepackage{graphicx}
\usepackage{dcolumn}
\usepackage{bm}
\usepackage{sidecap}
\usepackage{times}
\usepackage{verbatim}
\usepackage{graphicx}
\usepackage{color}
\usepackage{graphics}
\usepackage{amsmath}
\usepackage{float}

\usepackage{amssymb}
            
\usepackage[
            hyperindex,
            pdfstartview=FitH,
            plainpages=false]
            {hyperref}

\newcommand {\kbbf}{KBe$_2$BO$_3$F$_2$}

\newcommand {\uJcm}{$\mu$J/cm$^2$}

\begin{document}
\begin{CJK*}{GB}{gbsn}
\title{A time- and angle-resolved photoemission spectroscopy with probe photon energy up to 6.7 eV}
\author{Yuanyuan Yang}
\author{Tianwei Tang}
\author{Shaofeng Duan}
\author{Chaocheng Zhou}
\author{Duxing Hao}
\author{Wentao Zhang}
\email{wentaozhang@sjtu.edu.cn}
\affiliation{Key Laboratory of Artificial Structures and Quantum Control (Ministry of Education), School of Physics and Astronomy,
Shanghai Jiao Tong University, Shanghai 200240, China}
\affiliation{Collaborative Innovation Center of Advanced Microstructures, Nanjing 210093, China}
\date {\today}

\begin{abstract}

We present the development of a time- and angle-resolved photoemission spectroscopy based on a Yb-based femtosecond laser and a hemispherical electron analyzer. The energy of the pump photon is tunable between 1.4 and 1.9 eV, and the pulse duration is around 30 fs. We use a KBe$_2$BO$_3$F$_2$ non-linear optical crystal to generate probe pulses, of which the photon energy is up to 6.7 eV, and obtain an overall time resolution of 1 ps and energy resolution of 18 meV. In addition, $\beta$-BaB$_2$O$_4$ crystals are used to generate alternative probe pulses at 6.05 eV, giving an overall time resolution of 130 fs and energy resolution of 19 meV. We illustrate the performance of the system with representative data on several samples (Bi$_2$Se$_3$, YbCd$_2$Sb$_2$, FeSe).

\end{abstract}

\maketitle
\clearpage
\end{CJK*}

\section{Introduction}

Angle-resolved photoemission spectroscopy (ARPES) is a unique technique in probing the momentum-resolved electronic structure in solids\cite{Damascelli2003}, and has played an important role in revealing the electronic dispersions in high-temperature superconductors, topological insulators, graphene, and other quantum materials. In the past decade, the combination of ultrafast laser and ARPES extends such study to time realm, allowing detecting the nonequilibrium electronic dynamics in solids after ultrafast laser excitation, and such a technique is called time- and angle-resolved photoemission spectroscopy (trARPES). In trARPES experiments, electrons in occupied states are usually excited into unoccupied states without kicking electrons out of the material by an infrared femtosecond laser pulse (pump), and subsequently a femtosecond ultraviolet (UV) pulse (probe) photoemits electrons out of the material, then the energy and emission angle of the photoelectrons are analyzed by an electron analyzer. Time resolution is achieved by varying the time delay between the pump and probe pulses. TrARPES enables access into the study of ultrafast electronic dynamics, such as the unoccupied electronic states, relaxation of excited states by electron-electron or electron-phonon interactions, light-induced phase transitions, and so on, in superconductors\cite{Perfetti2007aa,Smallwood2012ab,Kummer2012aa,Rettig2012aa,Yang2014aa,Zhang2014aa,Rameau2014aa,Yang2015aa}, topological insulators\cite{Wang2012aa,Hajlaoui2013aa,Crepaldi2013aa,Sobota2014aa}, density wave systems\cite{Schmitt2008aa,Petersen2011aa,Rohwer2011aa,Liu2013aa,Ishizaka2011aa}, and other quantum materials\cite{Cavalieri2007aa,Papalazarou2012aa,Ulstrup2015aa,Tao2016aa,Sterzi2016aa}.

In trARPES experiments, it is a challenge to obtain suitable probe pulse. Firstly, to do photoemission, the probe photon energy must be higher than the work function of usual solid materials (usually $>$4 eV). Secondly, to get a reasonable time resolution, the pulse duration should be on the scale of sub-picoseconds. Thirdly, to reach a reasonable energy resolution in the study of low energy physics in quantum materials, the spectrum width in frequency of the probe should be on the scale of 10 meV. Lastly, to minimize the space charge effect in photoemission and also get reasonable count rates of the photoemitted electrons, the repetition rate of the laser pulse should be high enough (above kHz).

Usually, probe pulses in time-resolved ARPES can be divided into three classes according to the generation mechanism. The first class of the probe laser is based on the frequency conversion in nonlinear crystals, and the wavelength is in the UV range\cite{Carpene2009aa,Smallwood2012aa,Wang2012aa,Faure2012aa,Ishida2014aa,Yang2015aa}. The overall time resolution of such setups can be better than 100 fs and the corresponding energy resolution is on a scale of 10 meV. In such setups, $\beta$-barium borate (BBO) nonlinear crystals are usually used to do the frequency conversion, and the obtained photon energy is $<$6.3 eV. With such photon energy, the momentum range of solid materials that can be accessed is very limited in photoemission experiments. The other class of the probe pulse is generated through high-harmonic generations (HHG) process in gas phases. Such HHG probe pulse usually has much shorter wavelength (scale of 10 nm) and shorter pulse duration  ($<$50 fs) than that from nonlinear process in crystals\cite{Siffalovic2001aa,Nugent2002aa,Mathias2007aa,Dakovski2010aa,Wernet2011aa,Frietsch2013aa,Wang2015aa,Conta2016aa}. However, such probe pulse is usually broad in frequency due to the Fourier transform limit and has dramatic space charge effect in photoemission due to the low repetition rate, making it difficult to achieve high energy resolution in trARPES experiment. Recently, a high repetition rate (250 kHz) HHG system is developed in trARPES experiment but with very low probe photon flux because of the limited third-order harmonic generation efficiency\cite{Cilento2016aa}. A new class of probe laser under developing is in conjunction with high repetition rate x-ray free-electron laser sources\cite{Pietzsch2008aa,Hellmann2012aa,Oloff2014aa}. Such laser sources will have advantages of tunable photon energy and pulse duration in trARPES experiments. 

In this paper, we introduce the development of an advanced trARPES system with probe photon energy up to 6.7 eV in Shanghai Jiao Tong University. The system is based on a Yb-based laser, \kbbf~(KBBF) and BBO nonlinear optical crystals, and a hemispherical electron analyzer. KBBF nonlinear crystal, which pushes the frequency conversion to vacuum ultraviolet region (up to 7.5 eV), has been successfully used in realizing ultrahigh energy resolution (better than 1 meV) in ARPES experiment\cite{Liu2008}. Taking advantage of the KBBF nonlinear crystal, the probe photon energy of our trARPES is extended to 6.7 eV, which is capable of probing higher binding energy and larger momentum region than that probed by BBO-based 6 eV photon. The repetition rate of the laser pulse is tunable up to 500 kHz by 500/n, and the pump photon energy is tunable between 1.4 and 1.9 eV with pulse duration around 30 fs. Based on KBBF, we obtain an overall time resolution of 1 ps and energy resolution of 18 meV. In addition, alternative probe pulses with photon energy at 6.05 eV and tunable polarization based on BBO crystals give an overall time resolution of 130 fs and energy resolution of 19 meV. The two beams can be switched using a flipping mirror without replacing, adding, or removing any optics. The performance of the system is presented by measuring the ultrafast electronic dynamics in serval samples (Bi$_2$Se$_3$, YbCd$_2$Sb$_2$, FeSe ).

\section{Experimental setup}

\subsection{trARPES layout}
\begin{figure}
\centering\includegraphics[width=1\columnwidth]{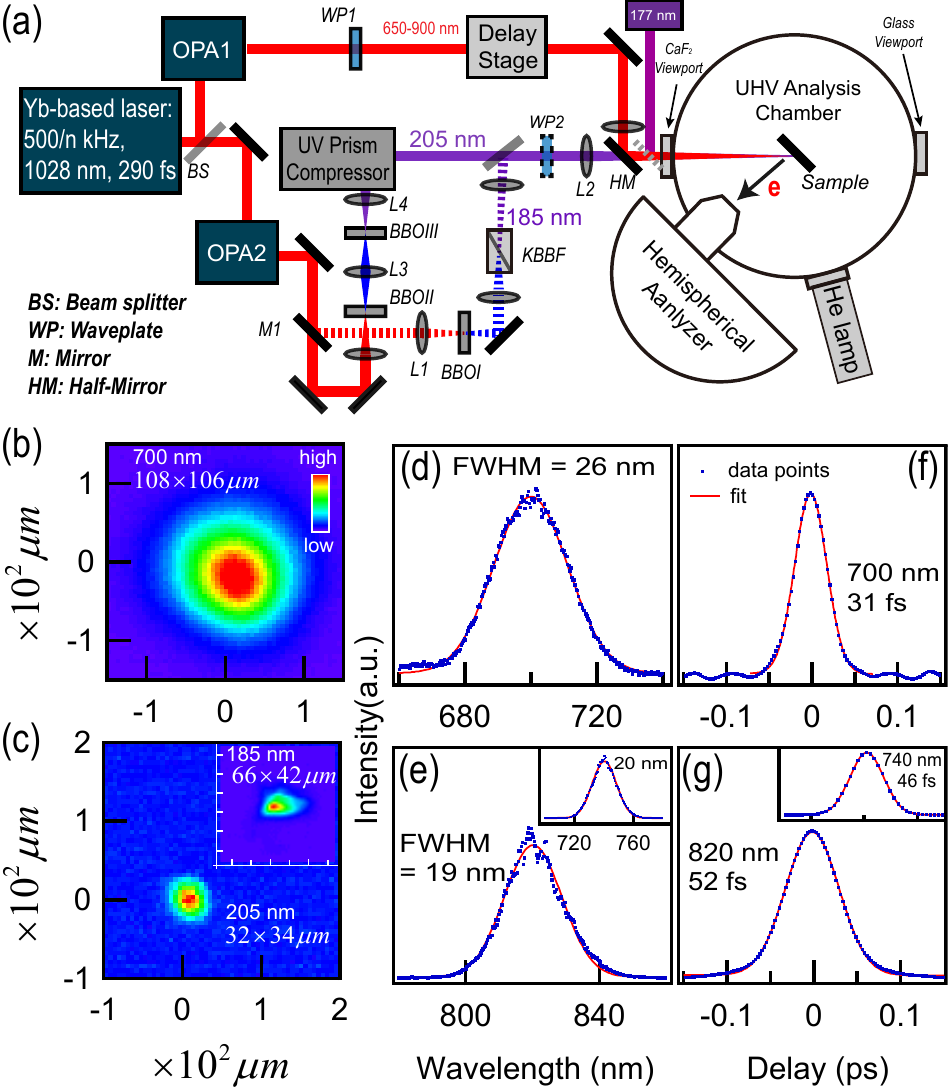}
\caption{
Experimental setup and characterizations of the laser beams and pulses. (a), Layout of the trARPES system.
(b) and (c), Beam profiles of pump (700 nm) and probe (205 nm) beams. The inset in (c) shows the beam profile of the 185 nm beam.
(d) and (e), Bandwidths of the pump and probe beams at fundamental frequencies.
(f) and (g), Autocorrelation signal of the pump and probe pulses at fundamental frequencies.
}
\label{Fig1}
\end{figure}

Figure~\ref{Fig1}(a) shows the layout of our trARPES setup, which includes a femtosecond laser system and an ARPES system. The fundamental ultrashort pulses of the laser system are generated by a commercial Yb-based amplifier. The output power is 15 W (at 500 kHz) with the wavelength centered at 1028 nm, the pulse duration is 290 fs, and the repetition rate is tunable at 500/n kHz. The output pulse are split into two with a ratio of $\sim$1:1 by a beam splitter (BS) and then coupled into two separate optical parametric amplifiers (OPA1, OPA2). The output wavelengths of the both OPAs are tunable between 650 nm and 900 nm with the pulse energy $>$1 $\mu$J and the pulse duration around 30 fs.

As shown in Fig.~\ref{Fig1}(a), the output of OPA1 is used as the pump beam directly, and the output wavelength is set to 700 nm in the performance tests in this paper. The polarization of the pump beam can be tuned freely using a half waveplate and a quarter waveplate (WP1), the working wavelengths of which are tunable. A motorized translation stage with step resolution of 1~$\mu$m is placed in the beam path of pump. Time resolution is achieved by varying the beam path difference between pump and probe using the translation stage. 

The output center wavelength of the OPA2 can be set to either 820 nm ($h\nu=1.5$ eV) or 740 nm ($h\nu=1.67$ eV) for fourth harmonic generation (FHG). The beam paths of 740 nm and 820 nm can be switched using the flipping mirror M1. When the M1 is on, the laser beam with the wavelength 740 nm is focused by the lens L1 onto a BBO crystal (BBOI) to generate second harmonic at 370 nm, and a KBBF device is placed right after the BBO to do the fourth harmonic generation. The output fourth harmonic at 185 nm is focused by the lens L2 into the ARPES chamber through a CaF$_2$ window. When the flipping mirror M1 is off, the laser beam with wavelength 820 nm is focused onto the BBOII to generate second harmonic at 410 nm, and then the focusing point is imaged by the lens L3 onto the BBOIII to do the fourth-harmonic generation. After collimated by the lens L4, the beam passes through a prism compressor, and then is focused by the lens L2 into the ARPES chamber. Before lens L2, optional tunable waveplate WP2 (a half waveplate or a quarter waveplate) is placed to vary the polarization of the probe beam.

To balance the energy resolution and time resolution, the output bandwidth of the OPA2 is set to $\sim$20 nm with the pulse duration of $\sim$50 fs.  Figures~\ref{Fig1}(d) and (e) show the spectra of the pulses at three different wavelengths generated by the two OPAs. We fit the spectra to one-dimensional Gaussians
\begin{equation}
\label{1D-Gaussian}
f(x)=f_\text{0}+C\cdot e^{-(\frac{x-x_\text{0}}{width})^2}
\end{equation}
to extract the $width$, and the corresponding FWHM of frequency is given by
\begin{equation}
\label{FWHM1}
FWHM=2\sqrt{\ln 2}\cdot width.
\end{equation}
The fitting results show the bandwidths at 700 nm, 740 nm, and 820 nm are 26 nm, 20 nm, and 19 nm, respectively.

The ARPES system includes five main parts. 1) A hemispherical electron analyzer (Scienta DA30L-8000R) is mounted on an analysis chamber, which has double magnetic shielding layers embedded, and the residue magnetic field at the focusing point of the analyzer is around 2 mGauss. 2) A sample manipulator, which has 5 translations and 1 rotation, is mounted on the top of the analysis chamber. The sample position and polar angle can be controlled precisely by the 3 motorized translations and 1 motorized rotation. A two-dimension hand-driven translation stage is mounted on top of a rotary seal to optimize the sample onto the rotary axis of the manipulator, and ensure the position of the beam spot on the sample is changeless when rotating the polar angle of the manipulator. With monitored by a high-resolution camera, the beam spot on the sample can be controlled with an accuracy less than 100 $\mu m$. The other two-dimension translation stage mounted below the rotary seal is used to optimize the sample onto the focusing point of the electron analyzer. 3) A loadlock system with two stages is used to load samples into the analysis chamber. 4) A helium discharge lamp (Scienta VUV5040) with a monochrometer (21 eV, 23 eV, and 41 eV) is mounted on the analysis chamber to characterize equilibrium electronic structures of sample. 5) A home-built compact 7 eV laser system based on the KBBF nonlinear crystal and a commercial picosecond laser is also coupled to the analysis chamber as an optional light source for regular ultrahigh energy resolution ARPES experiments.
 The analysis chamber and the second stage of the loadlock system are under UHV conditions with pressure better than $3\times10^{-11}$ Torr and $6\times10^{-11}$ Torr, separately. Using flowing liquid helium and a rotary vacuum pump to exhaust the outgoing helium gas and with a built-in resistive heater, the sample can be measured between 2 K and 500 K.

\subsection{Alignment of pump and probe beams}

A pinhole of 100 $\mu$m in diameter mounted at the bottom of the manipulator is used to do alignment. The pinhole is placed at the sample position when doing the alignment,  and a power meter is placed behind the glass viewport (see Fig.~\ref{Fig1}(a)) to measure the pump power. Alignment is done by tuning the half mirror in front of the CaF$_2$ viewport to make sure the pump beam go through the pinhole, and a good alignment is achieved when the power meter reads maximum. The probe beam is aligned by tuning an UV mirror to minimize the photoemission counts. The above processes ensure a good alignment for the pump and probe beams at the sample position.

The pump and probe beam profiles are characterized by a CCD beam profiler. A flipping mirror is placed in front of the CaF$_2$ viewport, and the CCD beam profiler is placed at the image position of the sample. In this way, the measured beam profiles are identical to that at the sample position. The measured beam profiles of pump and probe are shown in  Figs.~\ref{Fig1}(b) and (c), respectively. Generally, we firstly fit the beam profile to a two-dimensional Gaussian

\begin{equation}
\begin{split}
\label{2DGaussian}
f(x,y)=f_\text{0}+A\cdot e^{-\frac{1}{2(1-cor^2)}\cdot[(\frac{x-x_\text{0}}{x_\text{width}})^2}\\
^{+(\frac{y-y_\text{0}}{y_\text{width}})^2-\frac{2\cdot cor\cdot (x-x_\text{0})\cdot (y-y_\text{0})}{x_\text{width}\cdot y_\text{width}}]},
\end{split}
\end{equation}
and determine the rotation angle $\theta$ of the axes of the ellipsoidal beam spot related to horizontal direction by the relation
\begin{equation}
\begin{split}
\label{RAngle}
\theta=tan^{-1}(\frac{2\cdot x_\text{width}\cdot y_\text{width}}{y_\text{width}^2-x_\text{width}^2}\cdot cor),
\end{split}
\end{equation}
 and the FWHMs along the axis are given by

\begin{equation}
\begin{split}
\label{FWHM2}
\resizebox{1\hsize}{!}{$xFWHM= \sqrt{\frac{8\ln 2\cdot (1-cor^2)\cdot x_\text{width}^2\cdot y_\text{width}^2}{sin^2\theta\cdot x_\text{width}^2+cos^2\theta \cdot y_\text{width}^2+cor\cdot sin2\theta \cdot x_\text{width}\cdot y_\text{width}}},$}\\
\resizebox{1\hsize}{!}{$yFWHM= \sqrt{\frac{8\ln 2\cdot (1-cor^2)\cdot x_\text{width}^2\cdot y_\text{width}^2}{cos^2\theta\cdot x_\text{width}^2+sin^2\theta \cdot y_\text{width}^2-cor\cdot sin2\theta \cdot x_\text{width}\cdot y_\text{width}}}.$}
\end{split}
\end{equation}

The obtained spot size of pump beam is $108\times106~\mu$m, and the probe beams of wavelength at 205 nm and 185 nm are $32\times34~\mu$m and $66\times42~\mu$m, respectively. The above obtained values of pump and probe beam sizes are the normal incident case. Since the pulse energy of the pump can be higher than 1 $\mu J$, for a normal incidence, the pump fluence ($F$) of our setup can be higher than 3 mJ/cm$^2$ from the relation $F=\frac{4\ln2}{\pi}\cdot\frac{P}{xFWHM_\text{pump}\cdot yFWHM_\text{pump}+xFWHM_\text{probe}\cdot yFWHM_\text{probe}}$, in which $P$ is the pulse energy. Here since the angle between the pump and probe beam is $<$ 1$^\circ$, the effective projection of the two beam profiles is almost the same as that of the collinear case. For the general case of incidence, the pump fluence is calculated by $F\cdot cos(\theta)$, in which $\theta$ is the incident angle.

\section{Energy and time resolution}
 
\subsection{Time resolution}

In ultrafast experiments, time resolution is determined by both the pump and probe pulse durations. Positive group delay dispersion (GDD) induced by dispersive optical elements such as lens, windows, and waveplates causes a short laser pulse to spread in time as a result of different frequency components of the pulse traveling at different velocities. The expansion of the pulse duration $F_o$ (FWHM) can be write as
\begin{equation}
\label{FWHM3}
F_o(\lambda,L,F_i)=F_i\cdot \sqrt{1+\frac{16\cdot (\ln 2)^2\cdot GDD(\lambda,L)^2}{F_i^4}},
\end{equation}
in which $L$ is the length of the optical elements, and $F_i$ is the initial pulse duration.
Simple and cost effective prisms compressors, which can introduce negative GDD with certain setup, are usually used to compensate the positive GDD in common femtosecond setups based on oscillators, amplifiers and OPAs\cite{Fork1984}.

The pump pulse duration is measured by an optical autocorrelator, and the pulse duration is optimized by tuning the length of a built-in prism compressor in OPA1, and eventually a pulse duration of 31 fs in the test is obtained at the sample position as shown in Fig.~\ref{Fig1}(f). Identical length of optical elements (all CaF$_2$ optics) as that between OPA1 and the sample is temporarily put between OPA1 and the autocorrelator during optimizing. We note that the FWHM$_\text{pulse}$ of the pulse duration has a relation with the FWHM of autocorrelation signal by $FWHM_\text{pulse}=\frac{FWHM}{\sqrt{2}}$. The durations of the fundamental pulses of the probe are also characterized by the same method as shown in Fig.~\ref{Fig1}(g) and the inset.

For the probe pulse at 185 nm, the output bandwidth in frequency is limited due to the phase-matching bandwidth for the KBBF crystal with a thickness of 1.4 mm. We use the same method to simulate the time shape of the harmonic pulses at 185 nm as report in Krylov et al.\cite{krylov1995second}, and the simulation shows that the output pulse duration is on a scale of picosecond. Further calculation shows that positive GDD induced by the optical elements between the KBBF crystal and the sample has little effect in expanding the pulse of 1 ps duration, and thus it is not necessary to introduce negative GDD to compensate the positive GDD here.

For the probe pulse at 205 nm, simulation shows that the output pulse duration right after the BBOIII is around 100 fs, and the optical elements (total length $\sim$20 mm of CaF$_2$ for all the lenses and window) between the BBOIII and sample will expand the pulse duration to longer than 150 fs. A home-built prism compressor with negative GDD is introduced after the BBOIII to compensate the positive GDD, and the compressing length is optimized by checking the overall time resolution of the trARPES, as discussed in the following.

\begin{figure}
\centering\includegraphics[width=1\columnwidth]{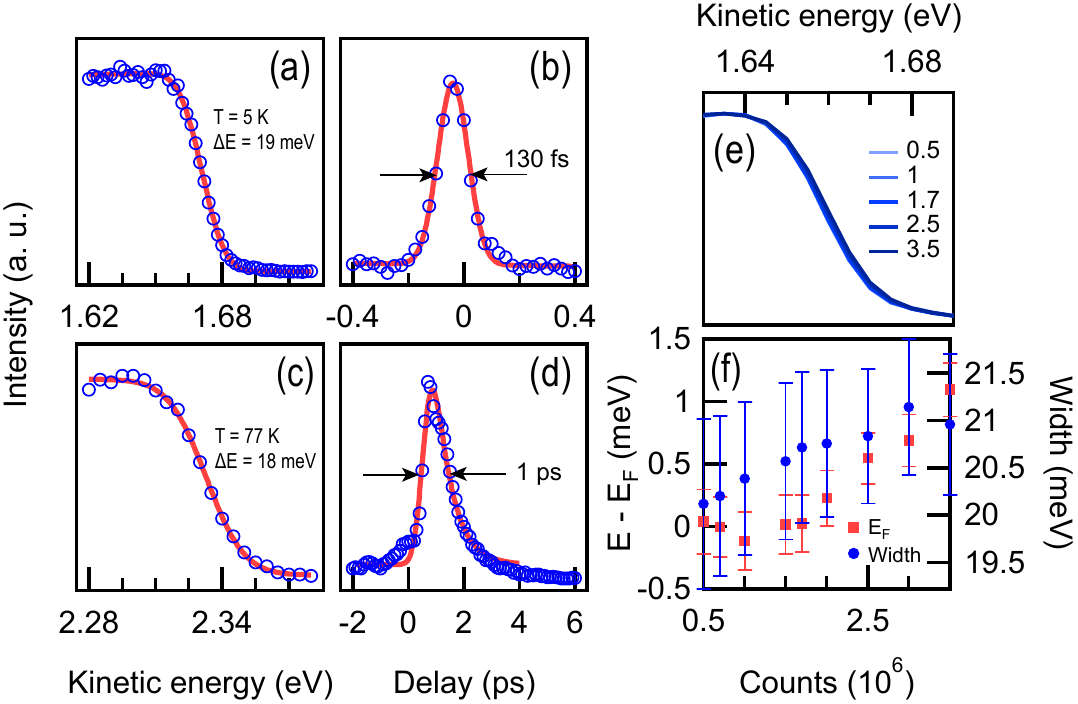}
\caption{
Energy and time resolution.
(a) and (b), Energy and time resolution of the system based on probe pulses at 205 nm measured on a polycrystalline gold at 5 K.
(c) and (d), Energy and time resolution of the system based on probe pulses at 185 nm tested on a Bi$_2$Se$_3$ single crystal at 77 K.
(e), Photoemission spectrum as a function of the total counts measured on a Bi$_2$Se$_3$ sample at 20 K.
(f), The position of the Fermi level and width of the Fermi edge as a function of photon flux.
}
\label{Fig2}
\end{figure}

It is a challenge to measure the probe pulse duration in UV range. An optical crosscorrelator is usually used to measure the duration of UV pulse, but the limited intensity of the probe beam here makes it difficult to do the cross correlation. Alternatively, we check the overall time resolution by doing trARPES on metallic samples, as reported by Smallwood et al.\cite{Smallwood2012aa} The time resolution was characterized by integrating the pump-probe photoemission signal far above the Fermi level ($E-E_F\geqslant1~$eV), where the lifetime of non-equilibrium quasiparticles is much shorter than the overall time resolution. For probes with central wavelength at 205 nm and 185 nm, we obtain upper limits of the time resolution R$_
\text{t}$ (FWHM) at 130 fs and 1 ps, respectively, as shown in  Figs.~\ref{Fig2}(b) and ~\ref{Fig2}(d). We note that the pulse duration  is 126 fs for the 205 nm probe pulse and 1 ps for the 185 nm pulse  by the relation $FWHM_{probe}=\sqrt{R_\text{t}^2-FWHM_{pump}^2}$.

\subsection{Energy resolution}
The overall energy resolution $R_\text{e}$ is the combination of the probe pulse bandwidth, the space charge effect induced broadening, and the resolution of the electron analyzer, and it can be extracted from measuring the Fermi edge of metallic samples at low temperature.  Figures~\ref{Fig2}(a) and ~\ref{Fig2}(c) show the photoemission spectroscopies on a poly crystalline gold sample at 5 K for probe beam at 205 nm, and on a Bi$_2$Se$_3$ single crystal at 77 K for probe beam at 185 nm. Function $I(\omega)$, which is the Fermi-Dirac distribution function $f(\omega)=\frac{1}{e^{\hbar\omega/k_\text{B}\cdot T}+1}$ convolved with a Gaussian resolution function $g(\omega)=e^{-\frac{4\ln2\cdot \omega^2}{R_\text{e}^2}}$,
\begin{equation}
\begin{split}
\label{FermiEdgeFunction}
I(\omega)=C_\text{0}+C_\text{1}\cdot\int_{-\infty}^{+\infty}\frac{1}{e^{\frac{\omega'-E_\text{F}}{k_\text{B}\cdot T}}+1}\\
\cdot e^{\frac{-4\ln2\cdot(\omega'-\omega)^2}{R_\text{e}^2}}d\omega',
\end{split}
\end{equation}
is used to fit the spectra near the Fermi edge to extract the energy resolution. The extracted overall energy resolution is 19 meV for the 205 nm probe, and 18 meV for the 185 nm probe.
We find that the time bandwidth product of the probe beam at 205 nm is $\sim$2400 meV$\cdot$fs, which is very close to the value $\sim$1800 meV$\cdot$fs of the Fourier transform limit for a Gauss-shape pulse, indicating that the probe optics are well designed and the home-built prism compressor works. With the energy resolution better than 20 meV, our trARPES system is capable of probing the ultrafast energy gap dynamics in superconductors, charge density wave materials, and other electronic excitations which need high energy resolution.

Photoemission with pulse light usually comes with space charge effect, which gives rise to energy shift and broadening in the spectrum\cite{ZHOU200527}. For ultrashort pulse light with pulse duration on the scale of 100 femtoseconds, the space charge effect is usually more pronounced than that of picosecond pulse. For a certain pulse duration, the space charge effect depends on the number of photoemitted electrons per pulse, and thus the photon energy is also important in the space charge effect, since for the same  flux higher energy photon usually photoemits more electrons due to the effect of inelastic scattering. We only tested the space charge effect for the 205 nm probe on a Bi$_2$Se$_3$ single crystal, because the space charge effect is weaker in principle for the 185 nm probe due to much longer pulse duration and larger spot size.

Since the Bi$_2$Se$_3$ sample is electrically grounded, we are not able to evaluate the number of photoemitted electrons per pulse by measuring the photocurrent. Alternatively, we measure the photoemission spectrum as a function of the counts persecond with pulse repetition rate at 500 kHz, energy fixed at 1.6 eV, 30 degree angular resolved mode, and pass energy 5 eV. The spectra is taken with the pass energy 5 eV, which is mostly used in our time-resolved experiments, and the result is shown in Fig.~\ref{Fig2}(e). Interestingly, we find barely any spectrum broadening or energy shift near the Fermi energy with increasing count rate, and detailed analysis shows that the Fermi level shifts only around 1 meV as the count rate increases from $5\times10^5$ to $3.5\times10^6$ (Fig.~\ref{Fig2}(f)). It is also worthy to mention that for the current detector and settings we use, the nonlinear effect of the electron analyzer near the Fermi energy is negligible\cite{Smallwood2012aa,Koralek2007}.

The above space charge effect test on a Bi$_2$Se$_3$ sample shows that for the photon energy at 6.05 eV it is possible to use high counts to take data on certain samples without apparent energy shift or band broadening. However, the above conclusion is only applicable on materials with high work function, in which fewer electrons can be photoemitted per pulse for a certain probe flux. Here the work function of the Bi$_2$Se$_3$ is $\sim$5.6 eV from the discussion in the following. Experimentally, space charge effect should be tested at the beginning of measuring any material and then a reasonable count rate can be determined to balance the efficiency of data collection and space charge effect. Moreover, for our setup the beam size of the probe beam is tunable by moving the focusing lens L2 towards to or away from the sample.

\section{Performance tests}

TrARPES tests are taken on a Bi$_2$Se$_3$ and a FeSe bulk single crystals for the probe photon energy at 6.05 eV, and on an YbCd$_2$Sb$_2$ single crystal for the probe photon energy at 6.7 eV. We select the pump photon at 700 nm with pulse duration $\sim$31 fs during the test. 

\subsection{TrARPES on Bi$_2$Se$_3$}

\begin{figure}
\centering\includegraphics[width=1\columnwidth]{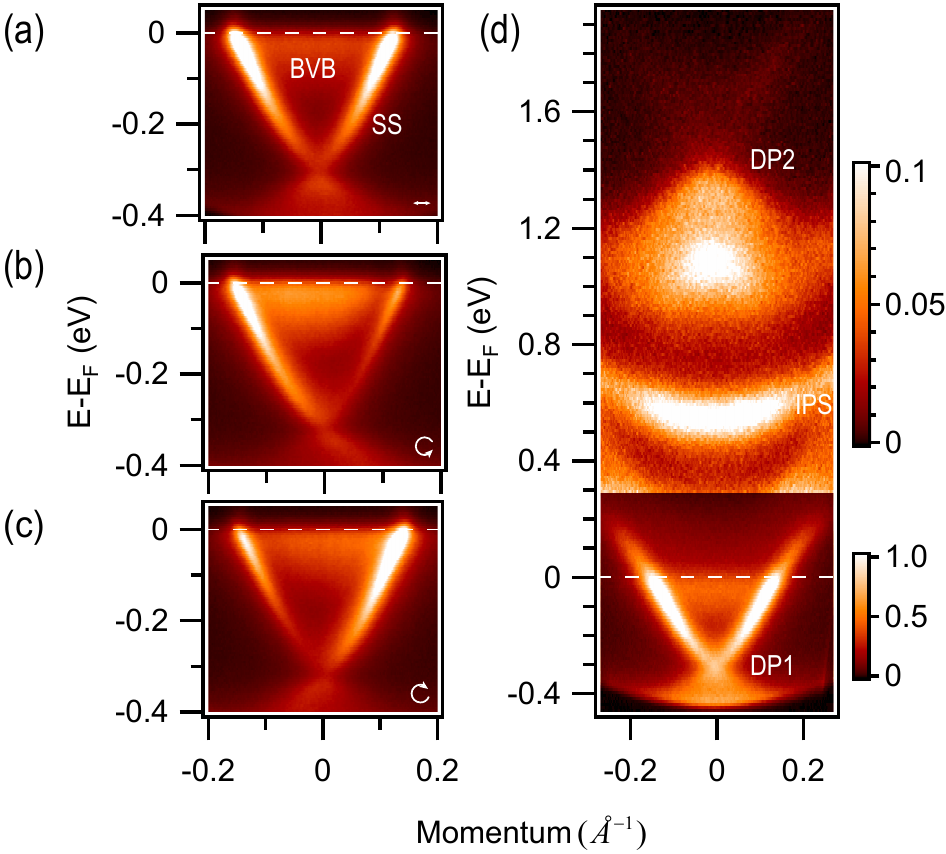}
\caption{
TrARPES test with 6.05 eV photon on Bi$_2$Se$_3$ at 77 K. 
(a), Equilibrium spectrum at negative delay time.
(b) and (c), Equilibrium spectra probed by left-circularly polarized and right-circularly polarized light.
(d), Time-resolved spectrum at the delay time 150 fs. The pump fluence is 43 \uJcm~with the repetition rate at 500 KHz.
}
\label{Fig3}
\end{figure}

The test of 6.05 eV probe is taken on a topological insulator Bi$_2$Se$_3$.  Figure~\ref{Fig3}(a) shows the equilibrium electronic states at 77 K, in which the Dirac point and also the surface state (SS) and the bulk conduction band (BCB) are clearly identified as reported in various publications. The data is taken with linear polarized probe beam. The energy of the photoemitted electrons at Fermi energy (E$_\text{F}$) in the material is 1.657 eV, which is set to 0 in the spectra, giving the system work function $\Phi=h\nu-E_\text{F}=4.4$ eV. Photoemission with circular dichroism in topological insulator is an indirect method to illustrate the spin polarization in energy bands. Figures~\ref{Fig3}(b) and (c) show the photoemission spectra with left-circular polarized light and right-circular polarized light, respectively. The intensity asymmetry and reversal after switching the light polarization indicate different spin polarization of the two bands\cite{Wang2012aa,Kondo_2017}. The observation indicates that our trARPES is capable of studying the ultrafast circular dichroism dynamics in solids.

The nonequilibrium electronic states measured at a delay time of 150 fs after the incidence of 700 nm pulse is shown in Fig. ~\ref{Fig3}(d). Firstly, we see a low energy cutoff ($E_\text{cutoff}$) at about -0.44 eV below the Fermi level. The cutoff is due to the work function barrier in photoemission, and the work function of Bi$_2$Se$_3$ we measured can be calculated by $W=h\nu-|E_\text{cutoff}|\sim5.61$ eV. Secondly, the most pronounced feature above the Fermi level is the first image potential state (IPS), which is the electronic state bound in front of the metallic Bi$_2$Se$_3$ surface. The IPS is populated before time-zero by 6.05 eV photon, probed by 1.77 eV photon, and decayed towards negative delays. The estimated bind energy of the IPS from the data is -0.8 eV related to the vacuum level, and this is consistent with the expected value of -0.85 eV from a metallic surface\cite{soifer2017band}. At last, except for the Dirac point (DP1) at -0.32 eV below the Fermi level, the pump probe spectra clearly shows another Dirac point (DP2) at energy 1.42 eV and the related linear dispersion far above the Fermi level, consistent with the DFT calculation\cite{soifer2017band}. The observations show the capability of our system to probe the unoccupied states far above the Fermi energy with high energy and time resolutions.

\subsection{TrARPES on FeSe}

\begin{figure}
\centering\includegraphics[width=1\columnwidth]{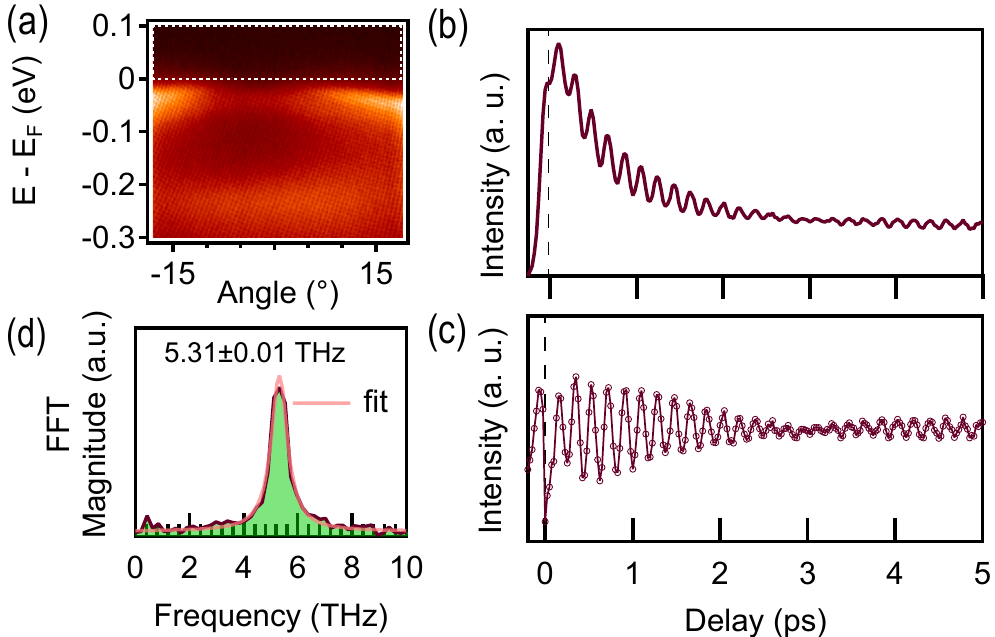}
\caption{
TrARPES on a bulk FeSe single crystal at 4 K.
(a), Equilibrium photoemission spectra at negative delay.
(b), The integrated intensity as a function of delay time.
(c), The intensity after removing the smooth delay background in (b). 
(d), FFT magnitude of the curves in (c).
 The pump fluence is 278 \uJcm~with the repetition rate at 500 KHz.
}
\label{Fig4}
\end{figure}

The performance of the high time resolution for the 6.05 eV probe is tested on a FeSe bulk sample, in which experimentally there is a hole-like pocket at the center of the Brillouin zone and electron-like pockets at the corners of the Brillouin zone below the structural transition temperature. Figure~\ref{Fig4}(a) shows the hole band near the gamma point probed by 6.05 eV photon. Figure~\ref{Fig4}(b) shows the photoemission intensity integrated in the window (dashed rectangle) shown in (a) as a function of delay time, in which an oscillation in intensity can be identified. Figure ~\ref{Fig4}(c) shows the intensity oscillation in time after removing the smooth decay background, and the corresponding Fourier transform magnitude as a function of frequency is shown in Fig.~\ref{Fig4}(d), giving an oscillation peaked at 5.31$\pm$0.01 THz ($\frac{1}{188~\text{fs}}$). Both the peak position and error are extracted from fitting the FFT curve to a Lorentzian. The oscillation at a period of 188 fs is consistent with the frequency of the oscillation due to locking in A$_\text{1g}$ phonon observed in the ultrafast x-ray diffraction\cite{gerber2017femtosecond}, but slightly higher than the 5.25 THz from a trARPES experiment at much higher pump fluence in the same paper. The observed oscillation with a period shorter than 200 fs demonstrates the ultrahigh time resolution of our trARPES system.

\subsection{TrARPES on YbCd$_2$Sb$_2$}

\begin{figure}
\centering\includegraphics[width=1\columnwidth]{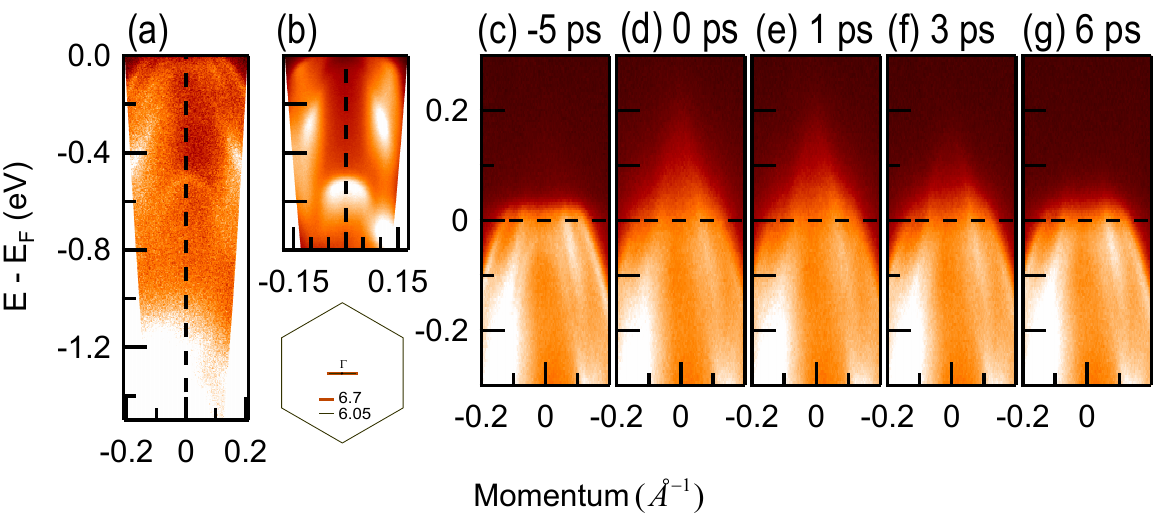}
\caption{
TrARPES on YbCd$_2$Sb$_2$ probed by 6.7 eV photons at 78 K. 
(a) and (b), equilibrium photoemission spectra probed by 6.7 eV and 6.05 eV photons, respectively. The momentum cuts are shown below (b) for the two photon energies.
(c)--(g), Time evolution of band structure tested near $\Gamma$ point of YbCd$_2$Sb$_2$ at various pump-probe delay. 
 The pump fluence is 71 \uJcm~with the repetition rate at 500 KHz.
}
\label{Fig5}
\end{figure}

The test of trAPRES based on KBBF for probe at 185 nm is taken on the ternary ytterbium transition-metal antimonide YbCd$_2$Sb$_2$. As shown in Figs.~\ref{Fig5}(a) and (b), with higher photon energy, electronic states with higher binding energies can be accessed, and the accessing range is at least 80$\%$ improved related to that of the 6.05 eV photon. We note that for the current setup of the electron analyzer only the counts of  photoemitted electrons with energy above 0.8 eV in the vacuum are reliable. Moreover, the momentum range that can be accessed at the Fermi energy for the same cut is improved by $\sim$17$\%$. Two hole-like pockets at the center ($\Gamma$ point) of the Brillouin zone in equilibrium can be identified in Fig.~\ref{Fig5}(c) at delay time -5 ps, but they are not clearly resolved by the 6.05 eV photon in (b). The difference between the two may be due to the matrix element effect or different $k_z$ probed in the k space. The time-resolved spectra are shown in Figs.~\ref{Fig5}(d)--(g). At delay 1 ps after photon excitation, the band top of the inner hole band at $\sim$0.2 eV above the Fermi level can be identified, and at the delay time 6 ps, the spectra almost recovers to its equilibrium state. TrARPES experiment with probe at higher photon energy of the 6.7 eV extends the capability to probe ultrafast electronic states at different $k_z$ with deeper energy, and in addition, with wider momentum region than that probed by the 6.05 eV photon.

\section{Conclusion}
In conclusion, we develop a time-resolved ARPES system based on a Yb-based laser and a hemispherical electron analyzer. The pump photon is tunable between 650 nm and 900 nm, and based on the BBO and KBBF nonlinear crystals, there are two options of probe photon energies (6.05 eV and 6.7 eV). The system can be running at 500/n kHz. The performance tests show an overall high-energy resolution of 19 meV and time resolution of 130 fs in trARPES based on the BBO nonlinear crystal, and shows a time bandwith product near the Fourier transform limit. We also show the success of applying the KBBF nonlinear crystal in trARPES with an overall time-resolution of 1 ps and energy resolution of 18 meV. The higher probe photon energy at 6.7 eV allows to probe the ultrafast electronic dynamics at different $k_z$ in momentum space with higher binding energy and wider momentum region. The successful application of KBBF in trARPES experiment indicates that it is possible to reach a high time resolution of 100 fs by developing KBBF crystals with thickness less than 0.1 mm to enhance the phase matching bandwidth, and push the probe photon energy up to 7.5 eV in the future.

\begin{acknowledgements}
We thank H. Wang for useful discussions and X. J. Zhou, L. Zhao, Y. H. Wang, and D. Qian for providing Bi$_2$Se$_3$ samples, J. Zhao for providing FeSe samples, and Y. F. Guo for providing YbCd$_2$Sb$_2$ samples. W.T.Z. acknowledges support from the Ministry of Science and Technology of China (2016YFA0300501) and from National Natural Science Foundation of China (11674224). T.W.T. acknowledges support from National Natural Science Foundation of China (11704247).
\end{acknowledgements}

\end{document}